\documentclass[10pt,twocolumn,conference]{IEEEtran}
\usepackage{dsfont}
\usepackage{setspace}
\usepackage{clrscode}
\usepackage{pict2e}

\usepackage{amsmath}
\usepackage{amssymb}
\usepackage[dvips]{graphicx}
\usepackage{epsfig}
\usepackage{algorithm}
\usepackage{algorithmic}
\usepackage{caption}
\usepackage{amsthm}
\usepackage{subcaption}

\usepackage[ps2pdf,
bookmarks=false,
bookmarksnumbered=false, 
bookmarksopen=false, 
colorlinks=false]{}

\makeatletter

\newcommand{\Rmnum}[1]{\expandafter\@slowromancap\romannumeral #1@}
\makeatother

\newcommand{\BS}{\text{BS}}
\newcommand{\UE}{\text{UE}}
\newcommand{\SINR}{\text{SINR}}

\newcommand{\g}{\text{g}}

\newcommand{\BE}{\text{BE}}

\newcommand{\Pout}{\text{P}_{\text{out}}}
\newcommand{\PoutL}{\text{P}_{\text{out},L}}
\newcommand{\PoutN}{\text{P}_{\text{out},N}}

\begin{document}
%
\title{Uplink Performance Analysis in D2D-Enabled mmWave Cellular Networks}

\author{Esma Turgut and M. Cenk Gursoy
\\Department of Electrical
Engineering and Computer Science, Syracuse University, Syracuse, NY, 13244
\\E-mail: eturgut@syr.edu, mcgursoy@syr.edu.}

\maketitle
\begin{abstract}
In this paper, we provide an analytical framework to analyze the uplink performance of device-to-device (D2D)-enabled millimeter wave (mmWave) cellular networks. Signal-to-interference-plus-noise ratio (SINR) outage probabilities are derived for both cellular and D2D links using tools from stochastic geometry. The distinguishing features of mmWave communications such as directional beamforming and having different path loss laws for line-of-sight (LOS) and non-line-of-sight (NLOS) links are incorporated into the outage analysis by employing a flexible mode selection scheme and Nakagami fading. Also, the effect of beamforming alignment errors on the outage probability is investigated to get insight on the performance in practical scenarios.
\end{abstract}

%

\thispagestyle{empty}


\section{Introduction}
Recent years have witnessed an overwhelming increase in mobile data traffic due to e.g., ever increasing use of smart phones, portable devices, and data-hungry multimedia applications. Limited available spectrum in microwave ($\mu$Wave) bands does not seem to be capable of meeting this demand in the near future, motivating the move to new frequency bands. Therefore, the use of large-bandwidth at millimeter wave (mmWave) frequency bands to provide much higher data rates and immense capacity has been proposed to be an important
part of the fifth generation (5G) cellular networks and has attracted considerable attention recently
\cite{Rappaport1} -- \cite{Ghosh}.

Despite the great potential of mmWave bands, they have been considered attractive only for short range-indoor communication due to increase in free-space path loss with increasing frequency, and poor penetration through solid materials. However, recent channel measurements and recent advances in RF integrated circuit design have motivated the use of these high frequencies for outdoor communication over a transmission range of about 150-200 meters \cite{Rappaport1}, \cite{Ghosh}. Also, with the employment of highly directional antennas, high propagation loss in the side lobes can be taken advantage of to support simultaneous communication with very limited or almost no interference to achieve lower link outage probabilities, much higher data rates and network capacity than those in $\mu$Wave networks.

Another promising solution to improve the network capacity is to enable device-to-device (D2D) communication in cellular networks. D2D communication allows proximity user equipments (UEs) to establish a direct communication link with each other
by bypassing the base station (BS). In other words, conventional two-hop cellular link is replaced by a direct D2D
link to enhance the network capacity. Network performance of D2D communication in cellular networks has recently been extensively studied as an important component of fourth generation (4G) cellular networks by using stochastic geometry, but it has been gaining even more importance in 5G networks and it is expected to be an essential part of mmWave 5G cellular networks.

Several recent studies have also addressed the mmWave D2D communication. In \cite{Qiao}, authors considered two types of D2D communication schemes in mmWave 5G cellular networks: local D2D and global D2D communications. Local D2D communication is performed by offloading the traffic from the BSs, while global D2D communication is established with multihop wireless transmissions via BSs between two wireless devices associated with different cells. The authors in \cite{Qiao} also proposed a resource sharing scheme to share network resources among local D2D and global D2D communications by considering the unique features of mmWave transmissions. In \cite{Guizani}, authors proposed a resource allocation scheme in mmWave frequency bands, which enables underlay D2D communications to improve the system throughput and the spectral efficiency. mmWave D2D multi-hop routing for multimedia applications was studied in \cite {Eshraghi} to maximize the sum video quality by taking into account the unique characteristics of the mmWave propagation.

Our main contributions can be summarized as follows:
\begin{itemize}
\item We provide an analytical framework to analyze the uplink performance of D2D-enabled mmWave cellular networks by using tools from stochastic geometry. In particular, we derive SINR outage probability expressions for both cellular and D2D links, considering different Nakagami fading parameters for LOS and NLOS components, employing the modified LOS ball model for blockage modeling, and considering a flexible mode selection scheme.

\item We investigate the effect of spectrum sharing type in SINR outage probability. Additionally, the effect of alignment errors on the SINR outage probability is investigated to get insight on the performance in practical scenarios.
\end{itemize}

\section{System Model} \label{sec:system_model}
In this section, the system model for D2D communication enabled mmWave cellular networks is presented. We consider a single-tier uplink network. BSs and UEs are spatially distributed according to two independent homogeneous Poison Point Processes (PPPs) $\Phi_{B}$ and $\Phi_U$ with densities $\lambda_{B}$ and $\lambda_U$, respectively, on the Euclidean plane. UEs are categorized as cellular UEs and potential D2D UEs with probabilities $q$ and $(1-q)$, respectively, where $q$ is the probability of being a cellular UE. A cellular UE is assumed to be associated with its closest BS. Potential D2D UEs have the capability of establishing a direct D2D link and can operate in one of the two modes according to the mode selection scheme: cellular and D2D mode. When operating in D2D mode, a UE can bypass the BS and communicate directly with its intended receiver. The density of UEs which communicate in D2D mode is $\lambda_d=(1-q)\lambda_U P_{D2D}$, and the density of UEs which communicate in cellular mode is equal to $\lambda_c=q\lambda_U+(1-q)\lambda_U (1-P_{D2D})$,
where $P_{D2D}$ is the probability of potential D2D UE selecting the D2D mode, and it will be described and characterized in detail later in the paper.

In this setting, we have the following assumptions regarding the system model of the D2D-enabled mmWave cellular network:

\textit{Assumption 1 (Directional beamforming):} Antenna arrays at the BSs and UEs are assumed to perform directional beamforming with the main lobe being directed towards the dominant propagation path while smaller side lobes direct energy in other directions. For tractability in the analysis, antenna arrays are approximated by a sectored antenna model \cite{Hunter}. The array gains are assumed to be constant $M_{\nu}$ for all angles in the main lobe and another smaller constant $m_{\nu}$ in the side lobe for $\nu \in \{\BS, \UE\}$. Initially, perfect beam alignment is assumed in between the transmitting nodes (e.g., cellular or potential D2D UEs) and receiving nodes (e.g., BSs or receiving D2D UEs) \footnote{Subsequently, beamsteering errors are also addressed.}, leading to an overall antenna gain of $M_{\BS}M_{\UE}$. Also, the beam direction of the interfering nodes is modeled as a uniform random variable on $[0,2\pi)$. Therefore, the effective antenna gain is a discrete random variable (RV) described by

\begin{equation}
    G=\left\{
                \begin{array}{ll}
                  M_{\BS}M_{\UE} &\text{w. p.} \; p_{M_{\BS}M_{\UE}}=\frac{\theta_{\BS}}{2\pi} \frac{\theta_{\UE}}{2\pi} \\
                  M_{\BS}m_{\UE} &\text{w. p.} \; p_{M_{\BS}M_{\UE}}=\frac{\theta_{\BS}}{2\pi} \frac{2\pi-\theta_{\UE}}{2\pi} \\
                  m_{\BS}M_{\UE} &\text{w. p.} \; p_{M_{\BS}M_{\UE}}=\frac{2\pi-\theta_{\BS}}{2\pi} \frac{\theta_{\UE}}{2\pi}  \\
                  m_{\BS}m_{\UE} &\text{w. p.} \; p_{M_{\BS}M_{\UE}}=\frac{2\pi-\theta_{\BS}}{2\pi} \frac{2\pi-\theta_{\UE}}{2\pi}
                \end{array}
              \right. \label{eq:antennagains}
\end{equation}
where $\theta_{\nu}$ is the beam width of the main lobe for $\nu \in \{\BS, \UE\}$, and $p_{G}$ is the probability of having a combined antenna gain of $G$.

\textit{Assumption 2 (Path-loss exponents and link distance modeling):} A transmitting UE can either have a line-of-sight (LOS) or non-line-of-sight (NLOS) link to the BS or the receiving UE. In a LOS state, UE should be visible to the receiving nodes, indicating that there is no blockage in the link. On the other hand, in a NLOS state, blockage occurs in the link. Consider an arbitrary link of length $r$, and define the LOS probability function $p(r)$ as the probability that the link is LOS. Using field measurements and stochastic blockage models, $p(r)$ can be modeled as $e^{-\zeta r}$ where decay rate $\zeta$ depends on the building parameter and density \cite{Bai1}. For simplicity, LOS probability function $p(r)$ can be approximated by a step function. In this approach, the irregular geometry of the LOS region is replaced with its equivalent LOS ball model. In this paper, modified LOS ball model is adopted similarly as in \cite{Andrews3}. According to this model, the LOS probability function of a link $p_L(r)$ is equal to some constant  $p_{L}$ if the link distance $r$ is less than ball radius $R_B$ and zero otherwise. The parameters $p_{L}$ and $R_B$ depend on geographical regions. $(p_{L,c}, R_{B,c})$ and $(p_{L,d}, R_{B,d})$ are the LOS ball model parameters for cellular and D2D links, respectively\footnote{Throughout the paper, subscripts $c$ and $d$ denote associations with cellular and D2D links, respectively.}. Therefore, LOS and NLOS probability function for each link can be expressed as follows:
\begin{align}\label{LOS_prob_funct}
  p_{L,\kappa}(r) &= p_{L,\kappa}\mathds{1}(r \le R_{B,\kappa}) \nonumber \\
  p_{N,\kappa}(r) &= (1-p_{L,\kappa})\mathds{1}(r \le R_{B,\kappa})+\mathds{1}(r > R_{B,\kappa})
\end{align}
for $\kappa \in \{c, d\}$ where $\mathds{1}(\cdot)$ is the indicator function. Different path loss laws are applied to LOS and NLOS links, thus $\alpha_{L,\kappa}$ and $\alpha_{N,\kappa}$ are the LOS  and NLOS path-loss exponents for $\kappa \in \{c, d\}$, respectively.

Since the link distance between D2D UEs is generally relatively small, we assume that the transmitting UEs are always LOS to the receiving UE, i.e., inside the LOS ball we have $p_{L,d}=1$, and therefore the path loss exponent for the D2D link is always equal to $\alpha_{L,d}$. For the sake of simplicity, we also assume that each potential D2D UE has its own receiving UE uniformly distributed within the LOS ball with radius $R_{B,d}$. Therefore, the probability density function (pdf) of the D2D link distance $r_d$ is given by $f_{r_d}(r_d)=2r_d/R_{B,d}^2$, $0 \leq r_d \leq R_{B,d} $. Pdf of the cellular link distance $r_c$ to the nearest LOS/NLOS BS is given by \cite{Bai2}
\begin{equation}\label{cellularlinkpdftonearestBS}
f_s(r_c)= 2\pi\lambda_B r_c p_{s,c}(r_c) e^{-2\pi\lambda_B\psi_s(r_c)}/\mathcal{B}_{s,c} \quad \text{for} \; s \in \{L,N\}
\end{equation}
where $\psi_s(r_c)=\int_0^{r_c} x p_s(x) dx$, $\mathcal{B}_{s,c}=1-e^{-2\pi\lambda_B\int_0^{\infty} x p_s(x) dx}$ is the probability that a UE has at least one LOS/NLOS BS, and  $p_s(x)$ is given in (\ref{LOS_prob_funct}) for $s \in \{L,N\}$. Similarly, given that the UE is associated with a LOS/NLOS BS, the pdf of the cellular link distance $r_c$ to the serving BS is
\begin{equation}
\hat{f}_s(r_c)= 2\pi\lambda_B r_c p_{s,c}(r_c) e^{-2\pi\lambda_B\big (\psi_s(r_c)+\psi_{s'}(r_c^{\alpha_{s,c}/\alpha_{s',c}})\big)}/\mathcal{A}_{s,c}
\label{cellularlinksdistancepdf}
\end{equation}
where $s \in \{L,N\}$, $s'$ is the complement of $s$, $\mathcal{A}_{s,c}$ denotes the association probability of a UE with a LOS and NLOS BS for $s=L$ and $s=N$, respectively. This probability is formulated as $\mathcal{A}_{L,c}=\int_0^{\infty} 2\pi\lambda_B r_c p_{L,c}(r_c) e^{-2\pi\lambda_B \big(\psi_L(r_c)+\psi_N(r_c^{\alpha_{L,c}/\alpha_{N,c}})\big)} dr_c$ for a LOS cellular link, and $\mathcal{A}_{N,c}=1-\mathcal{A}_{L,c}$ for a NLOS cellular link.

\subsection{Spectrum Sharing}
Cellular spectrum can be shared between cellular and D2D UEs in two different ways: underlay and overlay. In the underlay type of sharing, D2D UEs can opportunistically access the channel occupied by the cellular UEs. While for the overlay type of sharing, the uplink spectrum is divided into two orthogonal
portions, i.e.,  a fraction $\delta$ of the cellular spectrum is assigned to D2D mode and the remaining part $(1-\delta)$ is used for cellular communication, where $\delta$ is the spectrum partition factor \cite{Ghosh2}. Also, $\beta$ is defined as the spectrum sharing indicator which is equal to one for underlay and zero for overlay type of sharing.

\vspace{-0.06cm}
\subsection{Interference Modeling}
Each cellular UE is assigned a unique and orthogonal channel by its associated BS which means that there is no intra-cell interference between cellular UEs in the same cell. However, we assume universal frequency reuse across the entire cellular network causing inter-cell interference from the other cells' cellular UEs. In the underlay case, we focus on one uplink channel which is shared by the cellular and D2D UEs. Since the D2D UEs coexist with the cellular UEs in an uplink channel, they cause both intra-cell and inter-cell interference at the BSs and other D2D UEs. On the other hand, in the overlay case, since the uplink spectrum is divided into two orthogonal portions, there is no cross-mode interference, i.e., no interference from the cellular (D2D) UEs to the D2D (cellular) UEs. Moreover, we consider a congested network scenario in which density of cellular UEs is much higher than the density of BSs. Since $\lambda_U \gg \lambda_B$, each BS will always have at least one cellular UE to serve in the uplink channel. Therefore, the interfering cellular UEs in different cells is modeled as another PPP $\Phi_c$ with density $\lambda_B$. 

\subsection{Mode Selection}
In this work, a flexible mode selection scheme similarly as in \cite{Hesham1} is considered. In this scheme, a potential D2D UE chooses the D2D mode if the biased D2D link quality is at least as good as the cellular uplink quality. In other words, a potential D2D UE will operate in D2D mode if $T_d r_d^{-\alpha_{L,d}} \geq  r_c^{-\alpha_{s,c}}$, where $T_d \in [0, \infty)$ is the biasing factor, and $r_c$ and $r_d$ are the cellular and D2D link distances, respectively. Since we assume potential D2D UEs are always LOS to the receiving UEs, LOS path loss exponent $\alpha_{L,d}$ is used for the D2D links. Biasing factor $T_d$ has two extremes, $T_d=0$ and $T_d \to \infty$. In the first extreme case, D2D communication is disabled, while in the second case, each potential D2D UE is forced to select the D2D mode. The probability of selecting D2D mode, $P_{D2D}$, can be found as follows:

\begin{align}
P_{D2D}&=1-P_{cellular} \nonumber \\
&=1-\mathbb{E}_{r_d,r_c}\big [\mathbb{P}\{T_d r_d^{-\alpha_{L,d}} \leq  r_c^{-\alpha_{s,c}}\}\mathcal{B}_{s,c} \big ] \nonumber \\
&= 1-\mathbb{E}_{r_d,r_c}\bigg [\mathbb{P}\bigg \{r_c \leq r_d^{\alpha_{L,d} /\alpha_{s,c}}T_d^{-1/\alpha_{s,c}}\bigg \}\mathcal{B}_{s,c}\bigg] \nonumber \\
&=1-\hspace{-0.3cm}\sum_{s \in \{L,N\}} \int_0^{R_{B,d}} F_s\bigg(\frac{r_d^{a_s}}{T_d^{1/\alpha_{s,c}}}\bigg) f_{r_d}(r_d)\mathcal{B}_{s,c} dr_d \nonumber \\
&\stackrel{(a)}{=}1-\hspace{-0.3cm}\sum_{s \in \{L,N\}} \int_0^{R_{B,d}} \big(1-e^{-\pi\lambda_B \psi_s\big(\frac{r_d^{a_s}}{T_d^{1/\alpha_{s,c}}}\big)}\big) \frac{2r_d}{R_{B,d}^2} dr_d
\end{align}
where $a_s=\alpha_{L,d}/\alpha_{s,c}$, $F_s(r_c)=(1-e^{-\pi\lambda_B\psi_s(r_c)})/\mathcal{B}_{s,c}$ is the cumulative distribution function (cdf) of the cellular link distance $r_c$ to the nearest LOS/NLOS BS, and (a) follows from the substitution of the cdf of $r_c$ and pdf of $r_d$ into the expression.

\section{Analysis of Uplink SINR Outage Probability} \label{sec:Analysis of Uplink SINR Outage Probability}
In this section, we first develop a theoretical framework to analyze the uplink SINR outage probability for a generic UE using stochastic geometry. Although a biasing-based mode selection scheme is considered for selecting between D2D and cellular modes, the developed framework can also be applied for different mode selection schemes.
\subsection{SINR Analysis}
Without loss of generality, we consider a typical receiving node (BS or UE) located at the origin according to Slivyank's theorem for PPP. The SINR experienced  at a typical receiving node can be written as
\begin{equation}
SINR^{\kappa}=\frac{P_{\kappa}G_0h_0r_0^{-\alpha_{\kappa}(r_0)}}{\sigma^2+\underbrace{\sum_{i \in \Phi_c} P_cG_ih_ir_i^{-\alpha_{\kappa}(r_i)}}_{I_{c\kappa}}+\underbrace{\sum_{j \in \Phi_d} P_d G_jh_jr_j^{-\alpha_{\kappa}(r_j)}}_{I_{d\kappa}}}
\end{equation}
where $P_{\kappa}$ is the transmit power of the UE operating in mode $\kappa \in \{c,d\}$, $G_0$ is the effective antenna gain of the link which is assumed to be equal to $M_{\BS}M_{\UE}$, $h_0$ is the small-scale fading gain, $\alpha_{\kappa}(r_0)$ is the path-loss exponent of the link which is determined according to the LOS probability function, $r_0$ is the transmission distance, $\sigma^2$ is the variance of the additive white Gaussian noise component, $I_{c\kappa}$ is the aggregate interference at the receiving node from cellular UEs using the same uplink channel in different cells which constitute a PPP $\Phi_c$, and $I_{d\kappa}$ is the aggregate interference at the receiving node from D2D UEs located anywhere (hence including both inter-cell and intra-cell D2D UEs), which constitute another PPP $\Phi_d$. Actually, neither $\Phi_c$ nor $\Phi_d$ is a PPP due to the interaction between the point processes of BSs and UEs, and the mode selection scheme. Also, they are not independent. However, for analytical tractability based on the assumptions in \cite{Hesham1}, we assume interfering UEs operating in cellular mode and D2D mode constitute independent PPPs $\Phi_c$ and $\Phi_d$ with densities $\lambda_B$ and $\lambda_d$, respectively. A similar notation is used for $I_{c\kappa}$ and $I_{d\kappa}$, but note that the effective antenna gains $G_i$ and $G_j$, and path loss exponents $\alpha_{\kappa}(r_i)$ and $\alpha_{\kappa}(r_j)$ are different for different interfering links as described in (\ref{eq:antennagains}) and (\ref{LOS_prob_funct}), respectively. All links are assumed to be subject to independent Nakagami fading (i.e., small-scale fading gains have a gamma distribution). Parameters of Nakagami fading are $N_L$ and $N_N$ for LOS and NLOS links, respectively, and they are assumed to be positive integers for simplicity. When $N_L = N_N = 1$, Nakagami fading specializes to Rayleigh fading.

The above description implicitly assumes underlay spectrum sharing between cellular and D2D UEs. Note that since there is no cross-mode interference in the overlay case, the SINR expression in this case reduces to $SINR^{\kappa}=\frac{P_{\kappa}G_0h_0r_0^{-\alpha_{\kappa}(r_0)}}{\sigma^2+I_{\kappa \kappa}}$.

The uplink SINR outage probability $\Pout$ is defined as the probability that the received SINR is less than a certain threshold $\Gamma>0$, i.e., $\Pout= \mathbb{P}(\SINR<\Gamma)$. The outage probability for a typical UE in cellular mode is given in the following theorem.

\textit{Theorem 1:} In a single-tier D2D communication enabled mmWave cellular network, the outage probability for a typical cellular UE can be expressed as
\begin{align}
&\Pout^c(\Gamma)=\sum_{s \in \{L,N\}} \int_{0}^{\infty} \sum_{n=1}^{N_s} (-1)^{n} {N_s \choose n} e^{-\frac{n \eta_s \Gamma r_0^{\alpha_{s,c}}\sigma^2}{P_{c}G_0}} \times \nonumber \\
&\mathcal{L}_{I_{cc}}(\frac{n \eta_s \Gamma r_0^{\alpha_{s,c}}}{P_{c}G_0}) \mathcal{L}_{I_{dc}}(\frac{\beta n \eta_s \Gamma r_0^{\alpha_{s,c}}}{P_{c}G_0}) \hat{f}_{s}(r_0) \mathcal{A}_{s,c} dr_0
\end{align}
where
\begin{align}
&\mathcal{L}_{I_{cc}}(\frac{n \eta_s \Gamma r_0^{\alpha_{s,c}}}{P_{c}G_0})=\exp\bigg(-2\pi\lambda_B\bigg (\sum_{j \in \{L,N\}}  \sum_{i=1}^3 p_{G_i} \times \nonumber \\
&\bigg(\int_0^{\infty}\bigg(1-1/\bigg(1+\frac{n \eta_s \Gamma r_0^{\alpha_{s,c}}G_i}{G_0N_jt^{\alpha_{j,c}}}\bigg)^{N_j}\bigg)p_{j,c}(t)tdt\bigg)\bigg)\bigg)
\end{align}
and
\begin{align}
&\mathcal{L}_{I_{dc}}(\frac{\beta n \eta_s \Gamma r_0^{\alpha_{s,c}}}{P_{c}G_0})=\exp\bigg(-2\pi\lambda_d \bigg (\sum_{j \in \{L,N\}} \sum_{i=1}^3 p_{G_i} \times \nonumber \\
&\bigg(\int_0^{\infty}\bigg(1-1/\bigg(1+\frac{\beta n \eta_s \Gamma r_0^{\alpha_{s,c}}P_d G_i}{P_c G_0N_j t^{\alpha_{j,c}}}\bigg)^{N_j}\bigg)p_{j,c}(t)tdt\bigg)\bigg)\bigg)
\end{align}
are the Laplace transforms $\mathcal{L}_{I_{cc}}(v)$ and $\mathcal{L}_{I_{dc}}(\beta v)$ of $I_{cc}$ and $I_{dc}$ evaluated at $v=\frac{n \eta_s \Gamma r_0^{\alpha_{s,c}}}{P_{c}G_0}$, respectively, $\hat{f}_{s}(r_0)$ is the pdf of the cellular link distance given in (\ref{cellularlinksdistancepdf}), $\eta_s=N_s(N_s!)^{-\frac{1}{N_s}}$, and $p_{j,c}(\cdot)$ is given in (\ref{LOS_prob_funct}).

\textit{Proof:} The outage probability for a typical UE in cellular mode can be calculated as follows
\begin{align}
&\Pout^c(\Gamma)= \PoutL^c(\Gamma)\mathcal{A}_{L,c}+\PoutN^c(\Gamma)\mathcal{A}_{N,c} \nonumber \\
&\Pout^c(\Gamma)=\sum_{s \in \{L,N\}}\mathbb{P}\bigg( \frac{P_c G_0h_0r_0^{-\alpha_{s,c}}}{\sigma^2+I_{cc}+I_{dc}} \leq \Gamma \bigg) \mathcal{A}_{s,c} \nonumber \\
&=\hspace{-0.5cm}\sum_{s \in \{L,N\}}\hspace{-0.2cm} \int_0^{\infty}\hspace{-0.2cm} \mathbb{P}\bigg (h_0 \leq \frac{\Gamma r_0^{\alpha_{s,c}}}{P_cG_0}(\sigma^2+I_{cc}+I_{dc}) |r_0 \bigg)\hat{f}_{s}(r_0) \mathcal{A}_{s,c} dr_0 \nonumber \\
&= \hspace{-0.5cm} \sum_{s \in \{L,N\}} \hspace{-0.2cm} \int_0^{\infty} \hspace{-0.2cm} \sum_{n=1}^{N_s} (-1)^{n} {N_s \choose n} e^{-v\sigma^2} \mathcal{L}_{I_{cc}}(v) \mathcal{L}_{I_{dc}}(\beta v) \hat{f}_{s}(r_0) \mathcal{A}_{s,c} dr_0 \label{OutageProbability}
\end{align}
where $v=\frac{n \eta_s \Gamma r_0^{\alpha_{s,c}}}{P_{c}G_0}$, and (\ref{OutageProbability}) is derived noting that $h_0$ is a normalized gamma random variable with parameter $N_s$, and using similar steps as in \cite{Bai2}.

We can apply concepts from stochastic geometry to compute the Laplace transform of $I_{cc}$ and $I_{dc}$. The thinning property of PPP can be employed to split the $I_{\kappa c}$ into 6 independent PPPs as follows \cite{Bai3}:
\begin{align}
I_{\kappa c} &= I_{\kappa c,L} + I_{\kappa c,N} \nonumber \\
&=\sum_{G \in \big\{ \substack {M_{\BS}M_{UE},M_{\BS}m_{UE},\\ m_{\BS}M_{UE},m_{\BS}m_{UE}}\big \}} \sum_{j \in \{L,N\}}I_{{\kappa c,s}}^{G},  \label{eq:6PPP}
\end{align}
where $I_{\kappa c,L}$ and $I_{\kappa c,N}$ are the aggregate LOS and NLOS interferences arising from the cellular UEs using the same uplink channel in different cells for $\kappa=c$ and D2D UEs in the same cell and out-of-cell for $\kappa=d$, and $I_{\kappa c,j}^{G}$ denotes the interference for $j \in \{L,N\}$ with random antenna gain $G$ defined in (\ref{eq:antennagains}). According to the thinning theorem, each independent PPP has a density of $\lambda_Bp_{G}$ for $\kappa=c$ and $\lambda_d p_G$ for $\kappa=d$ where $p_{G}$ is given in (\ref{eq:antennagains}) for each antenna gain $G$.

Inserting (\ref{eq:6PPP}) into the Laplace transform expression and using the definition of the Laplace transform yield
\begin{align}
\mathcal{L}_{I_{\kappa c}}(v)&= \mathbb{E}_{I_{\kappa c}}[e^{-vI_{\kappa c}}]=\mathbb{E}_{I_{\kappa c}}[e^{-v(I_{\kappa c,L}+I_{\kappa c,N})}] \nonumber \\
&\stackrel{(a)}{=}\mathbb{E}_{I_{\kappa c,L}}\big[e^{-v \sum_{G} I_{\kappa c,L}^{G}}\big] \times \mathbb{E}_{I_{\kappa c,N}}\big[{e^{-v \sum_{G} I_{\kappa c,N}^{G}}}\big] \nonumber \\
&= \prod_G \prod_j \mathbb{E}_{I_{\kappa c,j}^G}[ e^{-vI_{\kappa c,j}^G}], \label{eq:LT}
\end{align}
where $G \in \{M_{\BS}M_{UE},M_{\BS}m_{UE}, m_{\BS}M_{UE},m_{\BS}m_{UE}\}$, $j \in \{L,N\}$, $v=\frac{n \eta_s \Gamma r_0^{\alpha_{s,c}}}{P_{c}G_0}$, and (a) follows from the fact that  $I_{\kappa c,L}$ and $I_{\kappa c,N}$ are interferences generated from two independent thinned PPPs. Now, we can compute the Laplace transform for $I_{\kappa c,j}^G$ using stochastic geometry as follows:
\begin{align}
\mathbb{E}_{I_{\kappa c,j}^G}[ e^{-vI_{\kappa c,j}^G}]&= e^{-2\pi\lambda_{\kappa}p_{G} \int_{0}^{\infty}(1-\mathbb{E}_h [ e^{-v P_{\kappa}G h t^{-\alpha_{j,c}} }])p_{j,c}(t) t dt} \nonumber \\
&\hspace{-1cm} \stackrel{(a)}{=} e^{-2\pi\lambda_{\kappa}p_{G} \int_{0}^{\infty} (1-1/(1+vP_{\kappa} Gt^{-\alpha_{j,c}}/N_j)^{N_j})p_{j,c}(t) t dt}, \label{eq:LT_LOS}
\end{align}
where $p_{j,c}(\cdot)$ is given in (\ref{LOS_prob_funct}), $\lambda_{\kappa}=\lambda_B$ for cellular interfering links and $\lambda_{\kappa}=\lambda_d$ for D2D interfering links. (a) is obtained by computing the moment generating function (MGF) of the gamma random variable $h$. By inserting (\ref{eq:LT_LOS}) into (\ref{eq:LT}), Laplace transform of $I_{\kappa c}$ can be obtained for $\kappa \in \{c,d\}$.

\textit{Theorem 2:} In a single-tier D2D communication enabled mmWave cellular network, the outage probability for a typical D2D UE can be expressed as
\begin{align}
&\Pout^d(\Gamma)=\int_{0}^{\infty} \sum_{n=1}^{N_L} (-1)^{n} {N_L \choose n} e^{-\frac{n \eta_L \Gamma r_0^{\alpha_{L,d}}\sigma^2}{P_{d}G_0}} \times \nonumber \\
&\mathcal{L}_{I_{dd}}(\frac{n \eta_L \Gamma r_0^{\alpha_{L,d}}}{P_{d}G_0}) \mathcal{L}_{I_{cd}}(\frac{\beta n \eta_L \Gamma r_0^{\alpha_{L,d}}}{P_{d}G_0}) f_{r_{d}}(r_0) dr_0
\end{align}
where
\begin{align}
&\mathcal{L}_{I_{dd}}(\frac{n \eta_L \Gamma r_0^{\alpha_{L,d}}}{P_{d}G_0})=\exp\bigg(-2\pi\lambda_d \bigg (\sum_{j \in \{L,N\}} \sum_{i=1}^3 p_{G_i} \times \nonumber \\
&\bigg(\int_0^{\infty}\bigg(1-1/\bigg(1+\frac{n \eta_s \Gamma r_0^{\alpha_{s,d}}G_i}{G_0N_jt^{\alpha_{j,d}}}\bigg)^{N_j}\bigg)p_{j,d}(t) tdt\bigg)\bigg)\bigg)
\end{align}
and
\begin{align}
&\mathcal{L}_{I_{cd}}(\frac{\beta n \eta_L \Gamma r_0^{\alpha_{L,d}}}{P_{d}G_0})=\exp\bigg(-2\pi\lambda_B \bigg (\sum_{j \in \{L,N\}} \sum_{i=1}^3 p_{G_i} \times \nonumber \\
&\bigg(\int_0^{\infty}\bigg(1-1/\bigg(1+\frac{\beta n \eta_s \Gamma r_0^{\alpha_{s,d}}P_d G_i }{P_d G_0N_jt^{\alpha_{j,d}}}\bigg)^{N_j}\bigg)p_{j,d}(t)tdt\bigg)\bigg)\bigg)
\end{align}
are the Laplace transforms $\mathcal{L}_{I_{dd}}(v)$ and $\mathcal{L}_{I_{cd}}(\beta v)$ of $I_{dd}$ and $I_{cd}$ evaluated at $v=\frac{n \eta_L \Gamma r_0^{\alpha_{L,d}}}{P_{d}G_0}$, respectively, $f_{r_{d}}(r_0)$ is the pdf of the D2D link distance given by $2r_d/R_{B,d}^2$ for $0 \leq r_d \leq R_{B,d}$, and $p_{j,d}(\cdot)$ is given in (\ref{LOS_prob_funct}).

\textit{Proof:} Proof follows similar steps as in the proof of Theorem 1, and the details are omitted for the sake of brevity.

\subsection{Uplink SINR Outage Probability Analysis In the Presence of Beamsteering Errors}
In Section \ref{sec:Analysis of Uplink SINR Outage Probability} and the preceding analysis, antenna arrays at the transmitting nodes (cellular or potential D2D UEs) and receiving nodes (BSs or UEs) are assumed to be aligned perfectly and uplink SINR outage probabilities are calculated in the absence of beamsteering errors. However, in practice, it may not be easy to have perfect alignment. Therefore, in this section, we investigate the effect of beamforming alignment errors on the outage probability analysis. We employ an error model similar to that in \cite{Wildman}. Let $|\epsilon|$ be the random absolute beamsteering error of the transmitting node toward the receiving node with zero-mean and bounded absolute error $|\epsilon|_{\text{max}} \le \pi$. Due to symmetry in the gain $G_0$, it is appropriate to consider the absolute beamsteering error. The PDF of the effective antenna gain $G_0$ with alignment error can be explicitly written as \cite{Marco2}
\begin{align}
\hspace{-0.5cm}f_{G_0}(\g)&=F_{|\epsilon|}\left(\frac{\theta_{\BS}}{2}\right)F_{|\epsilon|}\left(\frac{\theta_{\UE}}{2}\right)\delta(\g-M_{\BS}M_{\UE}) \nonumber \\
&+F_{|\epsilon|}\left(\frac{\theta_{\BS}}{2}\right)\left(1-F_{|\epsilon|}\left(\frac{\theta_{\UE}}{2}\right)\right)
\delta(\g-M_{\BS}m_{\UE}) \nonumber \\
&+\left(1-F_{|\epsilon|}\left(\frac{\theta_{\BS}}{2}\right)\right)F_{|\epsilon|}\left(\frac{\theta_{\UE}}{2}\right) \delta(\g-m_{\BS}M_{\UE}) \nonumber \\ &+\left(1-F_{|\epsilon|}\left(\frac{\theta_{\BS}}{2}\right)\right)\left(1-F_{|\epsilon|}\left(\frac{\theta_{\UE}}{2}\right)\right) \delta(\g-m_{\BS}m_{\UE}),
\label{eq:PDFofG}
\end{align}
where $\delta(\cdot)$ is the Kronecker's delta function, $F_{|\epsilon|}(x)$ is the CDF of the misalignment error and (\ref{eq:PDFofG}) follows from the definition of CDF, i.e., $F_{|\epsilon|}(x)=\mathbb{P}\{|\epsilon|\le x\}$. Assume that the error $\epsilon$ is Gaussian distributed, and therefore the absolute error $|\epsilon|$ follows a half normal distribution with $F_{|\epsilon|}(x)=\text{erf}(x/(\sqrt{2}\sigma_{\BE}))$, where $\text{erf}(\cdot)$ denotes the error function and $\sigma_{\BE}$ is the standard deviation of the Gaussian error $\epsilon$.

It is clear that all uplink SINR outage probability expressions in Section \ref{sec:Analysis of Uplink SINR Outage Probability} depend on the effective antenna gain $G_0$ between the transmitting and the receiving nodes. Thus, uplink SINR outage probability $\Pout^{\kappa}(\Gamma)$ for a typical UE in mode $\kappa \in \{c,d\}$ can be calculated by averaging over the distribution of $G_0$, $f_{G_0}(\g)$, as follows:
\begin{align}
\Pout^{\kappa}(\Gamma) &= \int_0^{\infty}\Pout^{\kappa}(\Gamma;g)f_{G_0}(\g)d \g \nonumber \\
&=F_{|\epsilon|}(\theta_{\BS}/2)F_{|\epsilon|}(\theta_{\UE}/2) \Pout^{\kappa}(\Gamma;M_{\BS}M_{\UE})+F_{|\epsilon|}(\theta_{\BS}/2) \nonumber \\
&\bar{F}_{|\epsilon|}(\theta_{\UE}/2) \Pout^{\kappa}(\Gamma;M_{\BS}m_{\UE})+ \bar{F}_{|\epsilon|}(\theta_{\BS}/2)F_{|\epsilon|}(\theta_{\UE}/2)  \nonumber \\ &\Pout^{\kappa}(\Gamma;m_{\BS}M_{\UE})+\bar{F}_{|\epsilon|}(\theta_{\BS}/2) \bar{F}_{|\epsilon|}(\theta_{\UE}/2) \Pout^{\kappa}(\Gamma;m_{\BS}m_{\UE}),
\end{align}
where we define $\bar{F}_{|\epsilon|}(\theta/2)=1-F_{|\epsilon|}(\theta/2)$.

\section{Simulation and Numerical Results}
In this section, theoretical expressions are evaluated numerically. We also provide simulation results to validate the the accuracy of the proposed model for the D2D-enabled uplink mmWave cellular network as well as to confirm the accuracy of the analytical characterizations. In the numerical evaluations and simulations, unless otherwise stated, the parameter values listed in Table \ref{Table} are used.

\begin{table}
\small
\caption{System Parameters}
\centering
  \begin{tabular}{| l | r|}
    \hline
    \textbf{Parameters} & \textbf{Values}  \\ \hline
    $\alpha_{L,c}$, $\alpha_{N,c}$; $\alpha_{L,d}$, $\alpha_{N,d}$ & 2, 4; 2, 4\\ \hline
    $N_{L}$, $N_{N}$ & 3, 2 \\ \hline
    $M_{\nu}$, $m_{\nu}$, $\theta_{\nu}$ for $\nu \in \{\BS, \UE\}$ & 20dB, -10dB, $30^o$  \\ \hline
    $\lambda_B$, $\lambda_U$, & $10^{-5}$, $10^{-3}$ $(1/m^2)$ \\ \hline
    $(p_{L,c}, R_{B,c})$, $(p_{L,d}, R_{B,d})$ & (1, 100), (1, 50) \\ \hline
    $q$, $\beta$, $\delta$, $T_d$ & 0.2, 1, 0.2, 1 \\ \hline
    $\Gamma$, $\sigma^2$ & 0dB, -74dBm \\ \hline
    $P_c$, $P_d$ & 200mW, 200mW  \\ \hline
  \end{tabular} \label{Table}
\end{table}

\begin{figure}
\centering
  \includegraphics[width=0.5\textwidth]{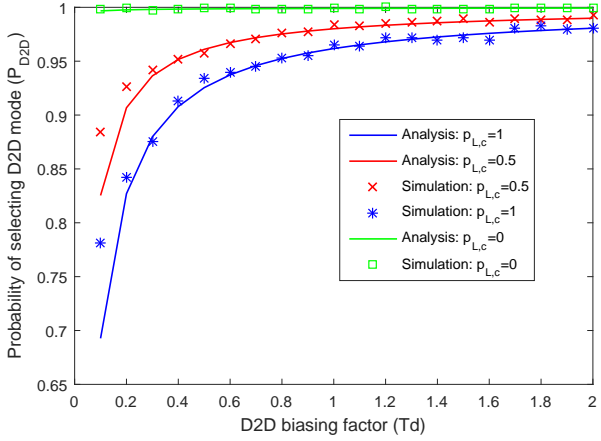}
  \caption{\small Probability of selecting D2D mode as a function of the D2D biasing factor $T_d$. \normalsize}
\label{Fig1}
\end{figure}

First, we investigate the effect of D2D biasing factor $T_d$ on the probability of selecting D2D mode for different values of LOS ball model parameter $p_{L,c}$ for the cellular link in Fig. \ref{Fig1}. As the D2D biasing factor increases, probability of selecting D2D mode expectedly increases. Also, since the number of LOS BSs increases with the increase in $p_{L,c}$, probability of selecting D2D mode decreases with increasing $p_{L,c}$.

\begin{figure}
\centering
  \includegraphics[width=0.5\textwidth]{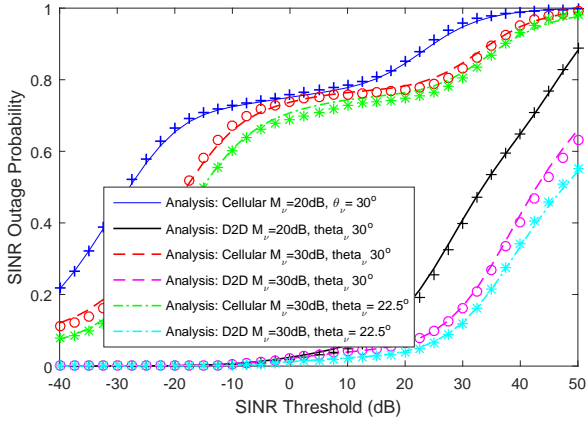}
  \caption{\small SINR outage probability as a function of the threshold in dB for different antenna parameters. Simulation results are also plotted with markers. \normalsize}
\label{Fig2}
\end{figure}

Next, we compare the SINR outage probabilities for different values of the antenna main lobe gain $M_{\nu}$ and beam width of the main lobe $\theta_{\nu}$ for $\nu \in \{\BS, \UE\}$ in Fig. \ref{Fig2}. Outage probability improves with the increase in the main lobe gain $M_{\nu}$ for the same value of $\theta_{\nu}$ for $\nu \in \{\BS, \UE\}$. Since we assume perfect beam alignment for serving links, outage probability increases with the increase in the beam width of the main lobe due to growing impact of the interference.

\begin{figure}
\centering
  \includegraphics[width=0.5\textwidth]{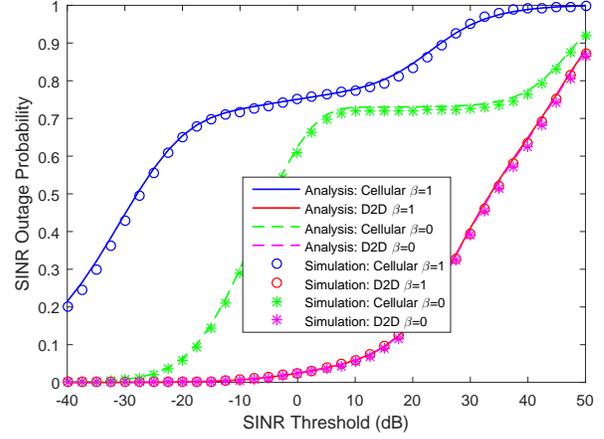}
  \caption{\small SINR outage probability as a function of the threshold in dB for different $\beta$ values. \normalsize}
\label{Fig3}
\end{figure}

In Fig. \ref{Fig3}, the effect of spectrum sharing type is investigated. As described in Section \ref{sec:system_model}, $\beta$ indicates the type of spectrum sharing; i.e., it is equal to one for underlay and zero for overlay scheme. For cellular UEs, outage probability is smaller in the overlay scheme compared to underlay since cross-mode interference from D2D UEs becomes zero in the case of overlay spectrum sharing. On the other hand, outage probability of D2D UEs remains same with both overlay and underlay sharing, showing that the effect of cross-mode interference from cellular UEs is negligible even under the congested network scenario assumption.

\begin{figure}
\centering
  \includegraphics[width=0.5\textwidth]{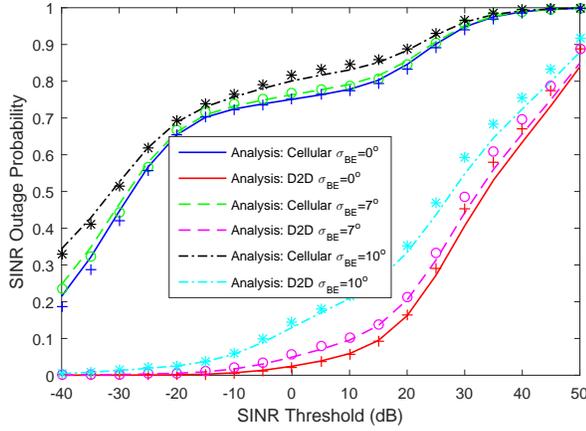}
  \caption{\small SINR outage probability as a function of the threshold in dB for different alignment errors
  $\sigma_{BE}$. Simulation results are also plotted with markers. \normalsize}
\label{Fig4}
\end{figure}

Finally, the effect of beam steering errors between the transmitting nodes (cellular or potential D2D UEs) and receiving nodes
(BSs or UEs) on the SINR outage probability of cellular and D2D links is shown  in Fig. \ref{Fig4}. As shown in the
figure, outage probability becomes worse for both cellular and D2D links with the increase in alignment error standard
deviation. Although the interference from interfering nodes remains unchanged, its effect grows with the increase in
alignment error on the main link. This proves the importance of having perfect beam alignment to achieve improved performance.

\section{Conclusion}
In this paper, we have provided an analytical framework to compute SINR outage probabilities for both cellular and D2D links in a D2D-enabled mmWave cellular network. Directional beamforming with sectored antenna model and modified LOS ball model for blockage modeling have been considered in the analysis. BSs and UEs are assumed to be distributed according to independent PPPs, and potential D2D UEs are allowed to choose cellular or D2D mode according to a flexible mode selection scheme. Numerical results show that probability of selecting D2D mode increases with increasing biasing factor $T_d$ and decreasing $p_{L,c}$. We have also shown that increasing the main lobe gain and decreasing the beam width of the main lobe result in lower SINR outage. Moreover, we have observed that the type of spectrum sharing plays a crucial role in SINR outage performance of cellular UEs. Finally, the effect of alignment error on outage probability is quantified. Analyzing  the link spectral efficiency of cellular and D2D UEs, and investigating the effect of using different mode selection schemes remains as future work.

\end{document}